\documentclass{article}
\usepackage{spconf,amsmath,graphicx,hyperref}
\usepackage{cite}
\usepackage{makecell}
\usepackage{amssymb,amsfonts}
\usepackage{float}
\usepackage{algorithmic}
\usepackage{textcomp}
\usepackage{xcolor}
\usepackage{booktabs}
\usepackage{enumitem}
\usepackage{hyperref}
\usepackage{xspace}
\hypersetup{
    colorlinks=true,  
    linkcolor=blue,   
    citecolor=blue,    
    filecolor=cyan,   
    urlcolor=magenta  
}
\hypersetup{breaklinks=true}
\usepackage[usestackEOL]{stackengine}

\newcommand{\etal}{\textit{et al}.\xspace}

\title{Generating Moving 3D Soundscapes with Latent Diffusion Models} 
\name{Christian Templin$^1$, Yanda Zhu$^2$ and Hao Wang$^1$}
\address{$^1$Stevens Institute of Technology, Hoboken, NJ, USA, \{ctemplin, hwang9\}@stevens.edu \\
         { $^2$Hunan Normal University, Changsha, China}, yandazhu87@hunnu.edu.cn}
\begin{document}
%
\maketitle
\begin{abstract}
Spatial audio has become central to immersive applications such as VR/AR, cinema, and music. Existing generative audio models are largely limited to mono or stereo formats and cannot capture the full 3D localization cues available in first-order Ambisonics (FOA). 
Recent FOA models extend text-to-audio generation but remain restricted to \textit{static} sources. 
In this work, we introduce \textit{SonicMotion}, the \textit{first} end-to-end latent diffusion framework capable of generating FOA audio with explicit control over moving sound sources. SonicMotion is implemented in two variations: 1) a descriptive model conditioned on natural language prompts, and 2) a parametric model conditioned on both text and spatial trajectory parameters for higher precision.
To support training and evaluation, we construct a new dataset of \textit{over one million} simulated FOA caption pairs that include both static and dynamic sources with annotated azimuth, elevation, and motion attributes. Experiments show that \textit{SonicMotion} achieves state-of-the-art semantic alignment and perceptual quality comparable to leading text-to-audio systems, while uniquely attaining low spatial localization error. 
\end{abstract}
\begin{keywords}Spatial Audio Generation, Latent Diffusion Model, Ambisonic Audio Generation
\end{keywords}

\section{Introduction}
\label{sec:intro}

The growing adoption of spatial audio has transformed the design of immersive media experiences in VR/AR, gaming, and music production. Among spatial formats, first-order Ambisonics (FOA) offers a compact 4-channel representation that encodes sound direction in three dimensions. While FOA is widely supported in consumer and professional platforms, creating FOA audio remains labor-intensive, requiring manual panning and expert software tools. This motivates the use of generative AI models to automate the production of spatial soundscapes directly from natural descriptions.

Recent advances in text-to-audio generation have demonstrated strong capabilities in producing coherent speech, sound effects, and music~\cite{pmlr-v202-liu23f,10.5555/3692070.3692575,2024arXiv240410301E}. However, these models typically generate \textit{mono} or \textit{stereo signals} and cannot express spatial attributes beyond left–right cues.
Several extensions have attempted to bridge this gap. For example, SpatialSonic~\cite{2024arXiv241010676S} generates stereo audio with controllable trajectories, but stereo lacks vertical and rear cues essential for immersion. 
Diff-SAGe~\cite{10888882} and ImmerseDiffusion~\cite{10889311} move closer by generating FOA audio, yet both are restricted to \textit{static sources} and therefore cannot represent \textit{moving objects} in 3D environments. Moreover, no publicly available dataset exists for training FOA models with dynamic motion.

\begin{figure*}[!t]
    \centering
    \includegraphics[width=6.5in]{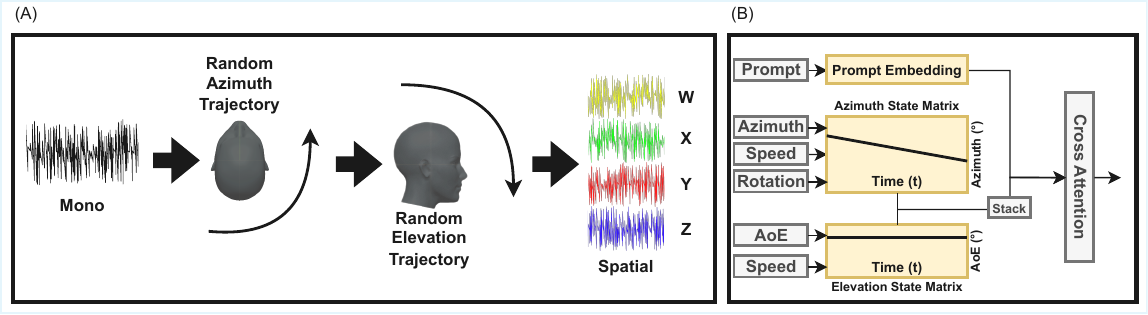} 
    \vspace{-0.2in}
    \caption{(A) Overview of dataset spatial augmentation pipeline. (B) \textit{SonicMotion} position conditioner's workflow.}
    \vspace{-0.2in}
    \label{fig:workflow}
\end{figure*}

This paper addresses these limitations through three key contributions:



    
\begin{itemize}[itemsep=0pt, parsep=0pt, topsep=0pt, partopsep=0pt]
    \item \textbf{\textit{SonicMotion}}, a novel generative framework that addresses these gaps by enabling FOA audio generation with controllable moving sound sources. To our knowledge, \textit{SonicMotion} is the first latent diffusion model capable of synthesizing 3D (Ambisonic) soundscapes with user-specified source trajectories. 
    \item \textbf{\textit{A new dataset}} of over 1M spatial audio-caption pairs, derived from existing captioned audio corpora and augmented with both static and dynamic FOA encodings. Each sample includes detailed metadata such as start/end angles, speed, and motion direction, enabling reproducible evaluation of spatial accuracy.
    \item Our experiments demonstrate that \textit{SonicMotion} matches the semantic quality of leading text-to-audio systems while providing substantially lower localization error for moving sources (e.g., azimuth and elevation errors reduced to within 15°). 
    Demonstration of our work is available at our website.\footnote{https://intellisys.haow.us/spatial-audio-project/} 
\end{itemize}

\vspace{-0.2in}
\section{Methodology}
\label{sec:method}

\vspace{-0.1in}
\subsection{FOA Format}

Ambisonics encodes 3D sound direction in a compact four-channel WXYZ format. First-order Ambisonics (FOA) is widely used for its simplicity and adequate spatial resolution: W is omnidirectional (pressure), and X, Y, Z capture directional components along the front–back, left–right, and up–down axes. 
In Equation Set~\eqref{eq1}, \(p\) represents the sound pressure, i.e., a mono audio signal. \(\theta\) and \(\phi\) represent the azimuthal angle and angle of elevation respectively. 
\vspace{0.1in}
\begin{equation}  
\begin{aligned}
    W &= \frac{1}{\sqrt{2p}},  & Y &= p \sin(\theta) \cos(\phi), \\
    X &= p \cos(\theta) \cos(\phi),  & Z &= p \sin(\phi). 
\end{aligned} \label{eq1}
\end{equation}

\subsection{Dataset}
To our knowledge there are no publicly available datasets consisting of spatial audio-caption pairs which include samples with moving sources. To overcome this lack of data, we introduce and release a dataset which involves spatially augmenting three existing audio-caption pair datasets. We select Clotho~\cite{9052990}, AudioCaps~\cite{kim-etal-2019-audiocaps}, and FreeSound~\cite{10.1145/2502081.2502245} as original datasets to be spatially augmented. First, we resample the audio to 16KHz, remove leading and trailing silences, and either truncate or loop the audio so that it is 10 seconds long. For each original audio sample, we generate one static and one dynamic spatial audio sample. To convert the mono audio signal to spatial audio we first randomly generate the initial azimuth and elevation angles to determine the direction. For spatial samples with a static source we can use these initial azimuth and elevation angles to directly encode the mono signal into FOA using Equation Set~\eqref{eq1}. For spatial samples which include a moving source we then randomly determine whether the azimuth, elevation, or both change over time. In any case, we randomly determine the final position of the angle such that it differs by at least $45^\circ$ for azimuth and $30^\circ$ for elevation. The speed of the moving source is randomly determined as ``fast,'' ``moderate,'' or ``slow.'' Movement start and end times are randomly determined based on the speed category. Finally, we linearly interpolate between initial and final angles in the movement window and encode into FOA using Equation Set~\eqref{eq1}. When interpolating between initial and azimuth angles, we randomly determine if the movement is clockwise or counter-clockwise. For each sample we record the initial and final azimuth and elevation angles, a clockwise flag, speed, and movement start and end times. 

Our dataset additionally includes a new set of captions which capture relevant details about the direction and movement of the sound source. To achieve this, we leverage Google's Gemma-3 12B \cite{gemma_2025} to automatically rephrase the original caption when provided with a list of relevant spatial parameters. Spatial parameters and their mappings to natural language are detailed in Table~\ref{tab:azimuth_mapping}. The final dataset consists of a training split with 1,018,430 samples, a validation split with 3,078 samples and a test split with 4,040 samples. 

\begin{table}[t]
\caption{Spatial Parameter Mapping}
\centering
    \scalebox{.8}{
        \begin{tabular}{llll}
        \toprule
        \textbf{Feature} & \textbf{Range} & \textbf{Bins} & \textbf{Map} \\
        \midrule
        Azimuth & $(-180.0^\circ, 180.0^\circ)$ &
        \makecell[l]{%
        $(-22.5^\circ, 22.5^\circ)$\\
        $(22.5^\circ, 67.5^\circ)$\\
        $(67.5^\circ, 112.5^\circ)$\\
        $(112.5^\circ, 157.5^\circ)$\\
        $(157.5^\circ, 180.0^\circ)$\\
        $(-180.0^\circ, -157.5^\circ)$\\
        $(-157.5^\circ, -112.5^\circ)$\\
        $(-112.5^\circ, -67.5^\circ)$\\
        $(-67.5^\circ, -22.5^\circ)$
        } &
        \makecell[l]{%
        front\\
        front-left\\
        left\\
        back-left\\
        back\\
        back\\
        back-right\\
        right\\
        front-right
        } \\
        \hline
        Elevation & $(-35.0^\circ, 35.0^\circ)$ &
        \makecell[l]{%
        $(-35.0^\circ, -30.0^\circ)$\\
        $(30.0^\circ, 35.0^\circ)$
        } &
        \makecell[l]{%
        down\\
        up
        }\\
        \hline
        Speed & (1.0s, 10.0s) &
        \makecell[l]{%
        (1.0s, 3.0s)\\
        (3.5s, 6.5s)\\
        (7.0s, 10.0s)
        } &
        \makecell[l]{%
        fast\\
        moderate\\
        slow
        } \\
        \bottomrule
        \end{tabular}
    }
    \vspace{-0.1in}
    \label{tab:azimuth_mapping}
\end{table}

\subsection{Spatial Autoencoder}
We use an autoencoder as a pretransform for compressing audio training data into latent space and decoding generated latents back into the original representation space. Autoencoders are frequently used to fulfill this need as their encoder and decoder can be decoupled to support both training and inference. We use the autoencoder provided by Descript Audio Codec (DAC)~\cite{10.5555/3666122.3667336} with some adaptations for FOA format. DAC is structured as a 1-dimensional U-Net autoencoder with a series of convolutions that repeatedly down-sample the input sequence during encoding and up-sample the encoded latents during decoding. To adapt DAC to our use case, we first replace the Residual Vector Quantization bottleneck with a continuous Variational Autoencoder bottleneck to avoid loss of information due to quantization which can degrade spatial performance. We remove the final \texttt{tanh()} activation from the decoder to prevent harmonic distortion~\cite{2024arXiv240410301E}. 
Next, we define a new loss function which better represents the FOA format. We use the loss function in Equation~\eqref{eq2}, which sums the Multi-Resolution Short Time Fourier Transform (MRSTFT) loss of each channel, the KL divergence loss of the bottleneck, as well as adversarial and feature matching losses applied to the discriminator. The \(\lambda\) parameter represents a weight which is perceptually updated during training, while the \(\beta\) parameter represents a weight which remains constant. The autoencoder compresses a 4-channel FOA signal into 64 channels with a down-sampling ratio of 2048 and an overall compression factor of 128. 
\begin{equation}
\begin{aligned}
    L_{\text{codec}} =  
    \frac{\lambda_{\text{mrstft}}}{4} L_{\text{mrstftW}}  + 
    \frac{\lambda_{\text{mrstft}}}{4} L_{\text{mrstftX}}  + 
    \frac{\lambda_{\text{mrstft}}}{4} L_{\text{mrstftY}}  + \\
    \frac{\lambda_{\text{mrstft}}}{4} L_{\text{mrstftZ}}   
    + \beta_{\text{kl}} L_{\text{kl}} 
    + \beta_{\text{adv}} L_{\text{adv}} 
    + \beta_{\text{fm}} L_{\text{fm}}. 
\end{aligned} \label{eq2}
\end{equation}

\subsection{Conditioning}
We propose two models which differ only in the extent of their conditioning. The descriptive model is conditioned on a text prompt embedding while the parametric model is conditioned on the same text embedding as well as a positional embedding which represents the azimuth and elevation of the sound at any time. We encode the text prompt using the pre-trained T5 encoder~\cite{10.5555/3455716.3455856}. As shown in an ablation study in Sun~\etal\cite{2024arXiv241010676S}, the spatial retrieval performance of the T5 encoder exceeds that of the CLAP \cite{10095969} text encoder as T5 is better at capturing spatial keywords in the prompt which are crucial for guiding the model in generating precisely defined spatial audio. To enhance the precision in the parametric model, we introduce an additional conditioner based on a state matrix that captures the position of the sound source with respect to time. We modify the position encoder from Sun~\etal\cite{2024arXiv241010676S} which provides guidance for azimuth based on the initial position, final position, and speed of the sound source. Modifications are made to support the FOA domain. Namely, we expand the range of azimuth to span a full 360$^\circ$ and account for both clockwise and counter-clockwise movement. We also accept additional parameters to encode the initial and final elevation. To build the state matrix, we first use Equation~\eqref{eq3} to describe the position of the sound, $\mu(t)$, given a current time, $t$, and duration, $T$. We build a zeros matrix of shape $(l,T)$ where $l$ is the number of bins representing unique positions and assign a value of 1 to represent the current position of the sound according to Equation~\eqref{eq4}. The guidance matrix is generated for both azimuth and elevation and these two matrices are concatenated. The resulting matrix is fused with the embedding of the text prompt using a cross-attention module with 4 layers and 4 heads. Our model also contains two additional cross-attention conditioners to include the temporal conditions, specifically start time and total time. 
\begin{equation}
    \mu(t) = \mu_{\text{start}} + \frac{t}{T} (\mu_{\text{end}} - \mu_{\text{start}}), 
    \label{eq3}
\end{equation}
\begin{equation}
    S_{\text{fine}}(l, T) =
    \begin{cases}
    1 & \text{if } l = | \mu(t) |, \\
    0 & \text{otherwise}.
    \end{cases} \label{eq4}
\end{equation}

\subsection{Diffusion Transformer}
We use the continuous diffusion transformer from Evans~\etal\cite{2024arXiv240410301E}. The model includes blocks with self-attention, cross-attention and multi-layer perceptron gates with layer normalization and skip connections. The conditioning token dimension is set to 768 and operates with a projection layer to fit all of the conditioners. We train with the objective of minimizing the Mean Squared Error between the ground truth latent and the predicted latent at the given timestep, for the given conditions. 

\subsection{Training}
The autoencoder is trained using randomly cropped 2 second segments of audio from the training split of the spatial dataset on a cluster of 4$\times$H100 GPUs. We use a batch size of 128 and train for 400K steps. The autoencoder generator uses an initial learning rate of $1e^{-4}$, while the discriminator uses an initial learning rate of $2e^{-4}$. Both the generator and discriminator use the AdamW optimizer with momentum values of 0.8 and 0.99 and a weight decay factor of $1e^{-3}$.

The diffusion model is similarly trained on a cluster of 4$\times$H100 GPUs. We use the full 10 seconds of audio from each sample in the training set. During training, we freeze the weights of the autoencoder and pre-trained T5 text encoder. We note that the weights of the cross attention mechanism in the position encoder remain unfrozen and are learned during training of the diffusion model. The model is trained with a batch size of 1024 for 100K steps. We use the AdamW optimizer with a learning rate of $1e^{-4}$, momentum values of 0.9 and 0.99 and a weight decay factor of $1e^{-3}$. The cross attention mechanism of the position encoder uses an increased learning rate of $2e^{-4}$. We additionally employ an inverse learning rate scheduler with an inverse gamma of $10^{6}$, power of 0.5 and warm up of 0.99. The descriptive model is trained using only spatial captions while the parametric model is trained for 50k steps using the original non-spatial captions and 50k steps using the spatial captions. 

\subsection{Evaluation}
\subsubsection{Autoencoder}
We evaluate the reconstruction quality of the autoencoder on the STFT and Mel Spectrogram (MEL) distance between the original audio and its reconstructed counterpart. These evaluation metrics are supported by the AuraLoss \cite{steinmetz2020auraloss} library and are implemented using the default settings. To evaluate the spatial accuracy of reconstructed audio we derive the intensity vectors for each cardinal direction by multiplying the omnidirectional W channel with the X, Y, or Z channel as shown in Equation Set~\eqref{eq5}. We then calculate the vectors representing azimuth ($\theta$) and elevation ($\phi$) at each sample in the audio using Equation Set~\eqref{eq6}. The error is computed as the L1 norm of the circular difference in azimuth and the linear difference in elevation between the original and reconstructed audio \cite{10889311}. 
\begin{equation}
    I_x = W \cdot X, \quad I_y = W \cdot Y, \quad I_z = W \cdot Z, 
    \label{eq5}
\end{equation}
\begin{equation}
    \theta = \tan^{-1} \left( \frac{I_y}{I_x} \right), \quad 
    \phi = \tan^{-1} \left( \frac{I_z}{I_x^2 + I_y^2} \right). 
    \label{eq6}
\end{equation}

We additionally report the error in direction of arrival (DoA) between original and reconstructed audio. This difference in spatial angle is calculated with Equation~\eqref{eq7} and Equation~\eqref{eq8} below:
\begin{equation}
    a = \sin^2\left( \frac{\Delta\phi}{2} \right) + \cos(\phi) \cdot \cos(\hat{\phi}) \cdot \sin^2\left( \frac{\Delta\theta}{2} \right), 
    \label{eq7}
\end{equation}
\begin{equation}
    \Delta_{\text{Spatial-Angle}} = 2 \cdot \arctan2\left( \sqrt{a}, \sqrt{1 - a} \right). 
    \label{eq8}
\end{equation}

\subsubsection{End-to-End Model}
The diffusion model is evaluated using CLAP score, Fréchet Distance (FD), Fréchet Audio Distance (FAD), and Kullback-Leibler Divergence (KL). To compute the CLAP score, we use a pretrained CLAP model to extract embeddings for the generated audio and its corresponding text prompt, then compute their cosine similarity. This measures the semantic alignment between audio and text. FD compares the distribution of CLAP audio embeddings between real and generated samples by computing the Fréchet Distance between their multivariate Gaussians, capturing semantic feature similarity. FAD compares VGGish \cite{45611} embeddings of real and generated audio, reflecting perceptual audio quality. KL divergence is computed between the distributions of CLAP audio embeddings for real and generated samples, quantifying divergence in their semantic representation distributions. Together, these metrics evaluate both the semantic accuracy and perceptual quality of the generated audio. As such, we extract and use only the omnidirectional W channel from our generated audio for these evaluations. When computing the CLAP score, we use the text embedding of the original non-spatial caption. To evaluate the spatial accuracy of generated audio, we utilize the same metrics as the autoencoder, namely, the difference in azimuth, elevation and spatial angle.

\section{Results and Discussions}

Evaluation results of the autoencoder are provided in Table~\ref{tab:autoencoder_metrics}. STFT and MEL reconstruction results prove that the autoencoder is effective at accurately encoding and decoding spatial audio. Evaluation of the spatial accuracy shows that the autoencoder near perfectly preserves the spatial attributes of the audio during the compression process. 

Evaluation results of our diffusion models are shown in Table~\ref{tab:model_metrics}. Comparison against state of the art text-to-audio models such as AudioLDM 2~\cite{10530074} and Stable Audio Open~\cite{10888461} reveals that both variations of our proposed model retain the ability to generate audio that aligns with the semantics of the sound source. It is important to note that this is strictly a comparison of the plausibility of generated audio and ignores any spatial attributes associated with audio generated from the proposed model. Thus, models are evaluated on audio generated using original non-spatial captions. We make this comparison to show that although our model generates a different format of audio than traditional models, it does not sacrifice any quality in the audio produced. 

Results of evaluating our proposed model on its spatial accuracy emphasizes the effectiveness of the additional conditioning parameters in our parametric model as it significantly outperforms its descriptive counterpart with a 51\% increase in overall angular performance. This can be attributed to the wide range of meaning when describing direction with natural language. For example, angles of $15^\circ$ to the left and $15^\circ$ to the right can both be reasonably described as ``front,'' despite being $30^\circ$ apart. While this error still falls within an acceptable range, the parametric model helps to avoid inherent ambiguity in natural language descriptions.

\begin{table}[t]
\caption{Comparison of proposed and existing models (SAO: Stable Audio Open, SM-D: \textit{SonicMotion} (Descriptive), SM-P: \textit{SonicMotion} (Parametric))}
\centering
\scalebox{.73}{\begin{tabular}{lccccccc}
\toprule
\textbf{Model} & \textbf{CLAP$\uparrow$} & \textbf{FD$\downarrow$} & \textbf{FAD$\downarrow$} & \textbf{KL$\downarrow$} & \textbf{L1($\theta$)$\downarrow$} & \textbf{L1($\phi$)$\downarrow$} & \textbf{$\Delta_{\text{angle}}$$\downarrow$} \\
\midrule
AudioLDM 2 & 0.22 & 0.55 & 1.78 & 0.27 & - & - & - \\
SAO & 0.19 & 1.18 & 7.90 & 0.59 & - & - & - \\
SM-D & 0.23 & 0.52 & 2.97 & 0.26 & 21.20$^\circ$ & 16.04$^\circ$ & 29.22$^\circ$ \\
SM-P & 0.23 & 0.57 & 3.00 & 0.28 & 13.17$^\circ$ & 4.01$^\circ$ & 14.32$^\circ$ \\
\bottomrule
\end{tabular}}
\vspace{-0.1in}
\label{tab:model_metrics}
\end{table}
\begin{table}[t]
\caption{Autoencoder evaluation results}
\centering
\scalebox{.8}{\begin{tabular}{lccccc}
\toprule
\textbf{Model} & \textbf{STFT$\downarrow$} & \textbf{MEL$\downarrow$} & \textbf{L1($\theta$)$\downarrow$} & \textbf{L1($\phi$)$\downarrow$} & \textbf{$\Delta_{\text{angle}}$$\downarrow$} \\
\midrule
\textit{SonicMotion} & 1.35 & 0.95 & 3.32$^\circ$ & 1.43$^\circ$ & 3.72$^\circ$ \\
\bottomrule
\end{tabular}}
\vspace{-0.1in}
\label{tab:autoencoder_metrics}
\end{table}

\section{Conclusion}

We presented \textit{SonicMotion}, the \textit{first} latent diffusion framework for first-order Ambisonics audio with controllable moving sound sources. Unlike prior text-to-audio or FOA models restricted to static positioning, \textit{SonicMotion} introduces two complementary approaches: a descriptive variant driven by natural language prompts and a parametric variant that integrates explicit spatial trajectories. Together with the release of a new large-scale dataset of over one million spatial audio–caption pairs—including detailed metadata on motion, speed, and direction—this work establishes a foundation for evaluating spatially dynamic sound generation at scale.
Our experiments show that \textit{SonicMotion} achieves state-of-the-art semantic alignment and perceptual quality while delivering substantial improvements in localization accuracy.

\bibliographystyle{IEEEbib}
\bibliography{references}

\end{document}